\documentclass{osa-article}

\journal{boe}
\articletype{Research Article}
\usepackage{lineno}
\usepackage{tikz}

\begin{document}

\title{In vivo three- and four-photon fluorescence microscopy using a 1.8~{\textmu}m femtosecond fiber laser system}

\author{Hideji Murakoshi,\authormark{1,2,*} Hiromi H. Ueda,\authormark{1,2} Ryuichiro Goto,\authormark{3} Kosuke Hamada,\authormark{4} Yutaro Nagasawa,\authormark{1,2} and Takao Fuji\authormark{4,$\dagger$}}

\address{
\authormark{1} Supportive Center for Brain Research, National Institute for Physiological Sciences, Okazaki, Aichi, 444-8585, Japan
\\
\authormark{2} Department of Physiological Sciences, The Graduate University for Advanced Studies, Hayama, Kanagawa, 240-0193, Japan
\\
\authormark{3} FiberLabs Inc., KDDI Laboratories Building, 2-1-15 Ohara, Fujimino, Saitama 356-8502, Japan
\\
\authormark{4} Laser Science Laboratory, Toyota Technological Institute, 2-12-1 Hisakata, Tempaku-ku, Nagoya, 468-8511, Japan
}
\email{\authormark{*}murakosh@nips.ac.jp, 
\authormark{$\dagger$}fuji@toyota-ti.ac.jp} %% email address is required


% The abstract should be limited to approximately 100 words.
\begin{abstract}
% We have demonstrated three- and four-photon fluorescence microscopy using a femtosecond fiber laser system which produces $\sim$150~fs micro joule pulses at 1.8~{\textmu}m. The system starts from a commercially available erbium doped silica fiber laser and the wavelength is converted to 1.8~{\textmu}m using a Raman shift fiber.  
% The 1.8~{\textmu}m pulses are amplified with a chirped pulse amplifier 
% based on a two-stage Tm:ZBLAN fiber amplifier. 
% The final output power is about 0.5~W at the repetition rate of 500~kHz. 
% The generated pulses are used for in vivo three- and four-photon fluorescence 
% microscopy. 
% We have clearly observed 0.8~mm deep inside the cortex of a living mouse.  

Multiphoton microscopy has enabled us to image cellular dynamics {\itshape in vivo}. 
However, the excitation wavelength for imaging with commercially available lasers is mostly limited between 650--1040~nm. 
Here we develop a femtosecond fiber laser system that produces $\sim$150~fs %microjoule 
pulses at 1.8~{\textmu}m. 
% We have demonstrated {\itshape in vivo} three/four-photon fluorescence microscopy using a femtosecond fiber laser system which produces $\sim$150~fs micro joule pulses at 1.8~{\textmu}m. 
Our 
%The 
system starts from an %commercially available 
erbium-doped silica fiber laser, 
and its wavelength is converted to 1.8~{\textmu}m using a Raman shift fiber. 
The 1.8~{\textmu}m pulses are amplified with %a chirped pulse amplifier based on 
a two-stage Tm:ZBLAN fiber amplifier. 
The final pulse energy is $\sim$1~{\textmu}J, 
% The final output power is 
%about 0.5~W at the repetition rate of 500~kHz, 
sufficient for {\itshape in vivo} imaging. 
% We demonstrate that the generated pulses can be used for {\itshape in-vivo} three/four-photon fluorescence microscopy. 
We successfully observe TurboFP635-expressing cortical neurons at a depth of 0.8~mm from the brain surface by three-photon excitation and Clover-expressing astrocytes at a depth of 0.2~mm by four-photon excitation.
\end{abstract}

%%%%%%%%%%%%%%%%%%%%%%%%%%  body  %%%%%%%%%%%%%%%%%%%%%%%%%%
\section{Introduction}
% For the last few decades multi-photon microscopy~\cite{BJ75-2015,NBio21-1369} has become very popular for brain science researchers 
% since it enables them to observe neurons in brains non-invasively in real-time. 
% Currently, Ti:sapphire mode-locked oscillators~\cite{OL16-42},   
% which are typically tunable from 650 to 1000~nm, 
% are most widely used femtosecond lasers 
% for two-photon microscopy and such microscopes based on a Ti:sapphire oscillator have already been commercially available. 
% By introducing green fluorescence proteins in neurons, 
% it has become possible to observe neurons non-invasively 
% up to typically a few hundred micrometer 
% deep inside mouse brains in real-time 
% using the two-photon microscope based on a Ti:sapphire oscillator. 
% However, the penetration depth is limited due to the strong light scattering in brains. 

In the last few decades, {\itshape in vivo} multiphoton microscopy~\cite{BJ75-2015,NBio21-1369} has become very popular in neuroscience. 
It enables non-invasive observation of fluorescent protein-expressing neurons up to $\sim$1000~{\textmu}m deep in mouse brains. 
While two-photon excitation is widely used, 3- or 4-photon excitation is rarely used. 
This is because the commercially-available femtosecond lasers, which only cover the wavelength between 650 and 1040~nm, and  
do not match the 3-photon (3P) /4-photon (4P) excitation of the commonly used fluorescent proteins such as green fluorescent protein (GFP) or Red fluorescent protein (RFP). e.g., 3P excitation of RFP and 4P excitation of GFP requires $>$1600~nm. 
Recently, custom-made lasers with wavelengths up to 1700~nm have been developed and applied for 3P/4P {\itshape in vivo} imaging~\cite{NMeth15-789,sciadv7-eabf3531}. 
However, the potential of 1.8~{\textmu}m for {\itshape in vivo} 3P/4P imaging has remained unknown.

% It would be possible to observe deeper inside brains 
% by using three photon fluorescence microscopy 
% because a longer wavelength laser 
% is used for three photon excitation of the fluorescence proteins~\cite{NPhoto7-205} and  
% the scattering of longer wavelength light is in general 
% lower than that of shorter wavelength. 
% Three- and four photon fluorescence microscopy 
% has already been realized 
% using optical parametric amplifiers pumped with 
% a high energy ($\sim$150~{\textmu}J) and high repetition rate  ($\sim$1~MHz) femtosecond ytterbium solid state amplifier~\cite{SciAdv7-eabf3531,eLife11-e63776}, 
% and Raman shift of the high energy ($\sim$6~{\textmu}J) femtosecond 1.5~{\textmu}m pulses from a powerful Er:fiber laser 
% using a rod style photonic crystal fiber~\cite{NPhoto7-205}.  

The light scattering of longer wavelengths in tissue is generally lower than that of shorter wavelengths. 
Therefore, 3-photon excitation with a longer wavelength would enable deeper imaging in brains over two-photon excitation~\cite{NPhoto7-205}. 
3P/4P fluorescence microscope at the wavelength of 1.3 and 1.7~{\textmu}m has already been realized using optical parametric amplifiers pumped with a high energy ($\sim$20~{\textmu}J) and high repetition rate ($\sim$1~MHz) femtosecond ytterbium solid state amplifier~\cite{sciadv7-eabf3531,eLife11-e63776}. 
In addition, Raman shift of the high energy ($\sim$6~{\textmu}J) femtosecond 1.5~{\textmu}m pulses from a powerful Er: fiber laser using a rod-style photonic crystal fiber~\cite{NPhoto7-205} has enabled 3P imaging.

% In this paper, we have demonstrated three- and four-photon fluorescence 
% microscopy using a fiber laser system consisting of rather standard components. 
% The system is being operated in a standard biomedical laboratory. 
% We were able to observe the whole cortex of a living mouse 
% using the system. 

In this paper, we show a new 1.8~{\textmu}m fiber laser system constructed with relatively standard components and operated in a standard biomedical laboratory. 
We demonstrate that the laser can be applied to the 3- /4-photon imaging of neuronal cells {\itshape in vivo}. 

\section{Laser system}

\begin{figure}
\centering\includegraphics[width=0.7\textwidth]{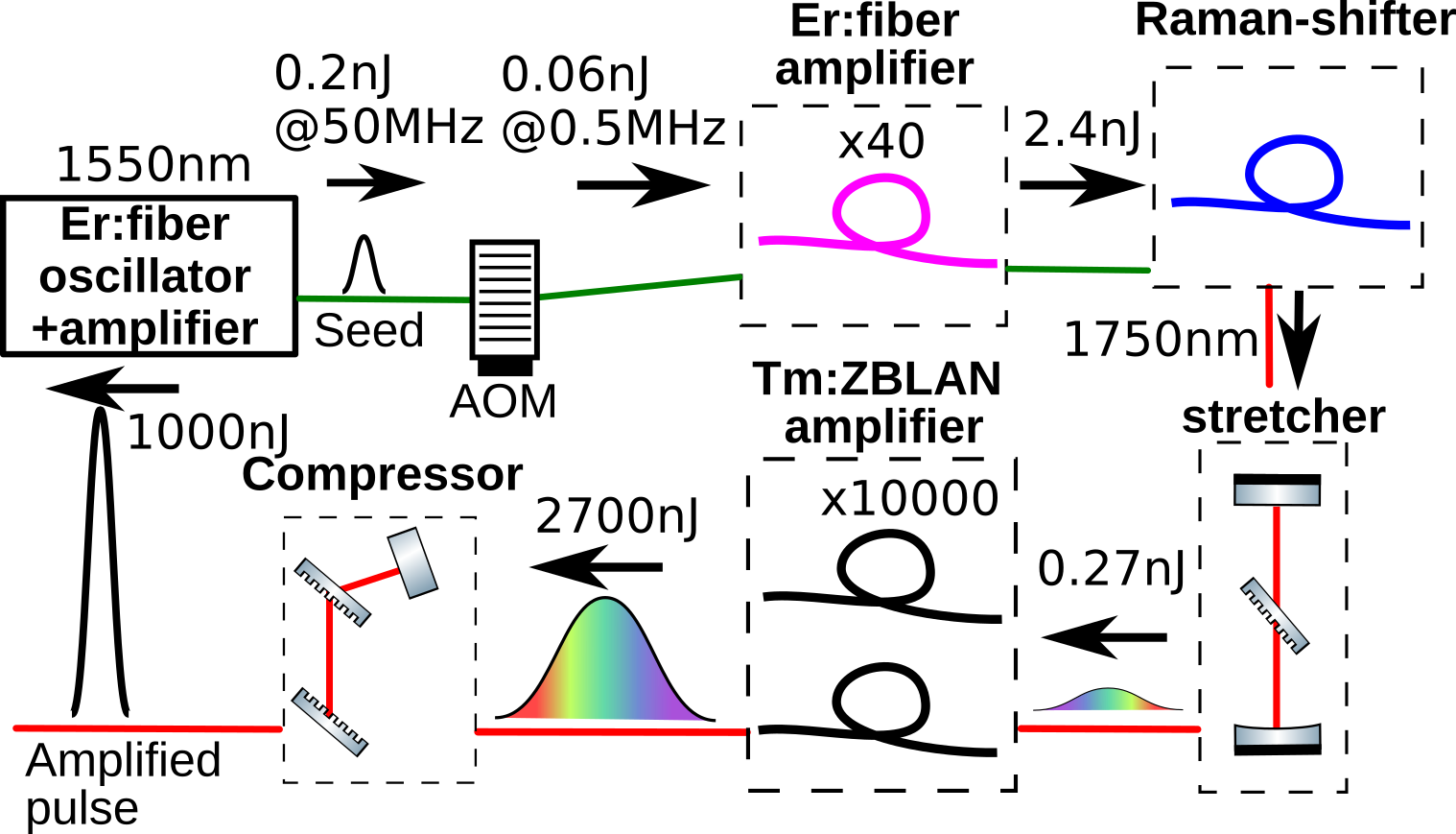}
\caption{\label{fig:schematic}Schematic of the laser system.}
\end{figure}

\begin{figure}
\begin{tikzpicture}
\draw (0,0) node [above right]{
\includegraphics[width=0.495\textwidth]{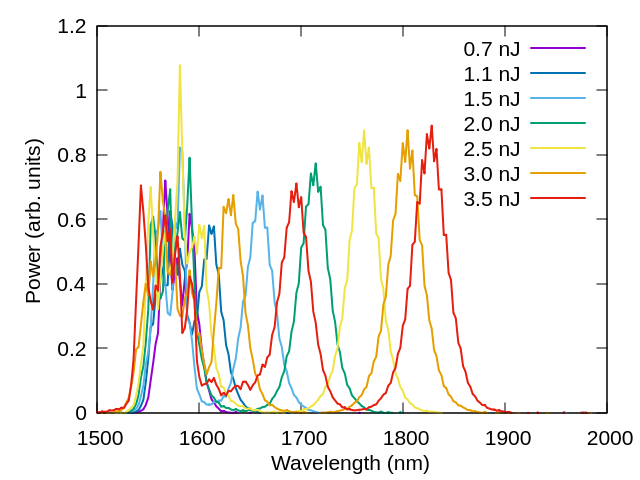}
\includegraphics[width=0.495\textwidth]{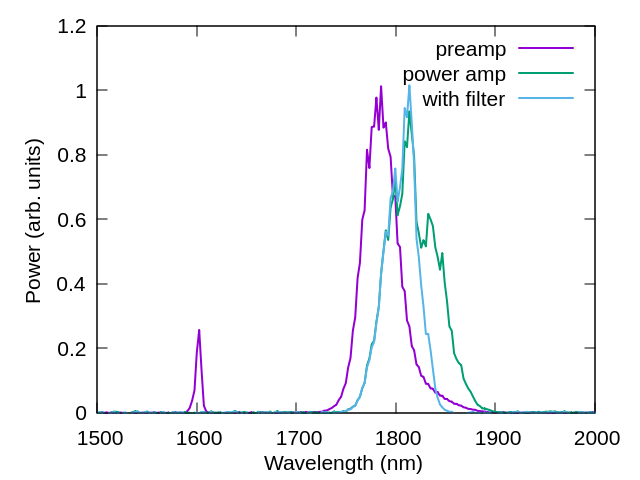}};
\draw (1.5,4.5) node [font=\small\sffamily]{(a)};
\draw (8.2,4.5) node [font=\small\sffamily]{(b)};
\end{tikzpicture}
\caption{\label{fig:spectra}Spectra of the output of (a) the Raman shift fiber, (b) Tm:ZBLAN preamplifier and power amplifier.}
\end{figure}

A schematic of our laser system is shown in Fig.~\ref{fig:schematic}. 
The system starts from a commercially available erbium-doped silica fiber (Er:SiO\textsubscript{2}) oscillator (VFLP-1560-M-fs, Connet). 
The output power of the oscillator and repetition rates are $\sim$1~mW and 50~MHz, respectively. 
The output of the oscillator was stretched with a dispersion compensation fiber with a length of 5~m (PMDCF, Thorlabs) 
and was amplified with a standard Er:SiO\textsubscript{2} amplifier (MFAS-ER-C-M-PA-PL, Connet). 
The repetition rate of the amplified pulse train was reduced from 50~MHz to 500~kHz by using an acousto-optic modulator (T-M300-0.1C16J-3-F2P, Gooch\&Hausego) driven by a pulse picker electronics and is amplified again with another same type Er:SiO\textsubscript{2} amplifier. 
The amplified pulse energy was about 7~nJ, 
and the pulses were sent to an anomalous dispersion fiber (PM1950) with a length of 4~m. 
The mode field diameter of the fiber is 8~{\textmu}m, 
and the estimated dispersion is $-$21.69~fs/mm\textsuperscript{2}. 
The pulse was compressed in the fiber, 
simultaneously inducing Raman scattering process inside the fiber. 
Figure~\ref{fig:spectra}(a) shows the Raman-shifted fiber's output spectrum as the amplifier's output pulse energy is varied. 
We were able to shift the center wavelength of the pulse up to 1.85~{\textmu}m. 
The center wavelength of the output laser pulse was set to 1.8~{\textmu}m and introduced into the next amplification stage. 
To generate the 1.8~{\textmu}m seed pulse, 
we used an all-fiber system consisting of lasers and optical components usually used for optical communications.

The 1.8~{\textmu}m pulse was sent to a Martines stretcher based on a transmission grating with grooves of 560~mm$^{-1}$ (PCG-560/2000-934, Ibsen). 
The stretcher introduces the group delay dispersion of $\sim$1.1~ps\textsuperscript{2} 
and stretches the pulse to $\sim$10~ps. 
The stretched pulse was sent to a thulium-doped fluoride fiber (Tm:ZBLAN) amplifier pumped by a continuous wave 1.6~{\textmu}m Er:SiO\textsubscript{2} laser (VFLS-1600-3W-PM, Connet) with a power of 2~W. 
The core diameter, the length, and the doping concentration of the fiber are 6~{\textmu}m, 30~cm, and 1~mol\%, respectively. 
The output spectrum of the preamplifier is shown in Fig.~\ref{fig:spectra}(b). 
The pulse was further amplified by another Tm:ZBLAN amplifier, which consists of a Tm:ZBLAN with a core diameter of 20~{\textmu}m, 
a doping concentration of 2~mol\% 
and a length of 20~cm. 
The pump laser is a commercially available Raman shift fiber laser (RLR-30-1620, IPG photonics). 
The amplifier's output power is 1.35~W at a pump power of 5.15~W, and the slope efficiency was 26.1\%. 

% The 1.8~{\textmu}m pulse is sent to a Martines stretcher based on a transmission grating with grooves of 560~mm$^{-1}$ (PCG-560/2000-934, Ibsen).  
% The stretcher introduces the group delay dispersion of $\sim$1.1~ps\textsuperscript{2} 
% and stretches the pulse to $\sim$10~ps.  
% The stretched pulse is sent to a thulium doped fluoride fiber (Tm:ZBLAN) amplifier pumped by a continuous wave 1.6~{\textmu}m Er:SiO\textsubscript{2} laser (VFLS-1600-3W-PM, Connet) with a power of 2~W. 
% The output spectrum of the preamplifier is shown in Fig.~\ref{fig:spectra}(b). 
% The pulse is further amplified by another Tm:ZBLAN amplifier 
% which consists of a Tm:ZBLAN with a core diameter of 20~{\textmu}m 
% , a doping concentration of 2~mol\%, and a length of 20~cm. 
% The pump laser is a commercially available Raman shift fiber laser 
% (RLR-30-1620, IPG photonics). 
% The output power of the amplifier is 1.35~W 
% at a pump power of 5.15~W.   
% The slope efficiency is 26.1\%. 

\begin{figure}
\begin{tikzpicture}
\draw (0,0) node [above right]{%
\includegraphics[scale=0.425]{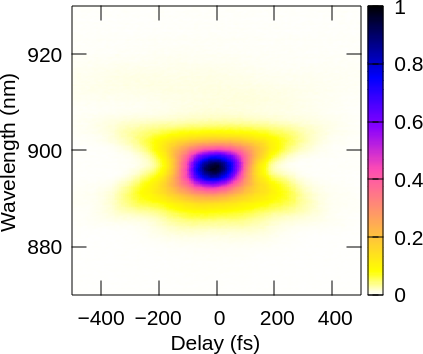}
\includegraphics[scale=0.33]{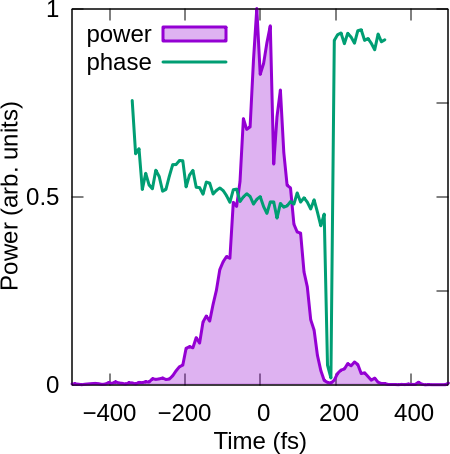}
\includegraphics[scale=0.33]{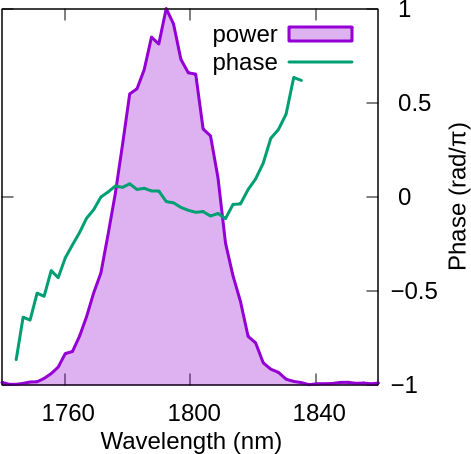}
};
\draw (1.5,3.7) node [font=\small\sffamily]{(a)};
\draw (8.6,3.7) node [font=\small\sffamily]{(b)};
\draw (9.2,3.7) node [font=\small\sffamily]{(c)};
\end{tikzpicture}
\caption{\label{fig:frog}(a) Measured FROG trace and 
retrieved power and phase in (b) time- and (c) frequency-domain. 
The FROG error is 0.29\%}
\end{figure}

The pulse was introduced into a Tracy compressor based on a pair of the same transmission gratings as that used in the stretcher. 
The distance of the pair of the gratings is 8.5~cm which corresponds to $-$0.77~ps\textsuperscript{2}. 
The spectrum after the compressor is shown in Fig.~\ref{fig:spectra}(b). 
The pulse was characterized by a second harmonic generation frequency-resolved optical gating (FROG) device (MS-FROG, FemtoEasy). The experimentally measured FROG trace and retrieved pulses in time- and frequency-domain are shown in Fig.~\ref{fig:frog}. There was some residual third-order dispersion (TOD) in the retrieved pulse because of the TOD mismatch between the dispersions of the fibers for the amplifier and the grating compressor. 
The pulse duration (full width at half maximum) was 142~fs. 

% The pulse is introduced into a Tracy compressor 
% based on a pair of the same transmission gratings 
% as that used in the stretcher. 
% The distance of the pair of the gratings 
% is 8.5~cm which corresponds to $-$0.77~ps\textsuperscript{2}. 
% The pulse is characterized 
% with a second harmonic generation 
% frequency resolved optical gating (FROG) device (MS-FROG, FemtoEasy). 
% The experimentally measured FROG trace 
% and retrieved pulses in time- and frequency-domain 
% are shown in Fig.~\ref{fig:frog}. 
% There is some residual third order dispersion (TOD) 
% in the retrieved pulse 
% because of the TOD mismatch between the dispersions 
% of the fibers for the amplifier 
% and the grating compressor. 
% The pulse duration is 142~fs. 

The retrieved spectrum does not have any intensity in the wavelength range longer than 1.84~{\textmu}m, 
although the spectrum recorded with a spectrometer has a significant intensity in the region. 
We believe that the long wavelength component is a sort of amplified spontaneous emission and is incoherent. 
Therefore, the component does not produce any second harmonic signals, 
which are supposed to show up in the FROG trace if it is coherent. 
The FROG trace did not change when we put a metallic plate in the grating compressor to block the long wavelength component.
The spectrum is also shown in Fig.~\ref{fig:spectra}(b). 
The long wavelength components might be absorbed and converted to heat in the water, damaging the brain tissue. 
Therefore, we kept blocking the long wavelength component and applied the output pulse for the multiphoton microscope. 
In the end, the power of the laser is 0.5~W in front of the microscope. 

% The retrieved spectrum does not have any intensity 
% in the wavelength range longer than 1.84~{\textmu}m 
% although the spectrum recorded with a spectrometer 
% has significant intensity in the region. 
% We believe that the long wavelength component 
% is a sort of amplified spontaneous emission 
% and is basically incoherent. 
% Therefore, the components do not produce 
% any second harmonic signal 
% which are supposed to show up in the FROG trace. 
% The FROG trace did not change 
% when we put a metalic plate in the grating compressor 
% to block the long wavelength component. 
% The spectrum is also shown in Fig.~\ref{fig:spectra}(b). 
% The long wavelength components might be absorbed 
% in the water in the sample 
% and make the damage threshold of the sample lower.
% We keep blocking the long wavelength components 
% and applied the output pulse for the multi-photon microscope. 
% In the end, the power of the laser is 0.5~W 
% in front of the microscope.

\section{Three-photon (3P) microscopy}

We built a laser scanning multiphoton microscope with our 1.8 µm laser as previously described~\cite{OSAC3-1428}. 
Briefly, the system consists of galvanometer scanners (6210H, Cambridge Technology), a scan lens (SL50-3P, Thorlabs), a trinocular tube (U-TR30IR, Olympus), and a water-immersion objective lens (XLPLNN25X SVVMP2SP, NA1.0, Olympus~\cite{OSAC3-1428}. 
The system was controlled with the ScanImage software~\cite{Pologruto_2003_BEO_2_13}. The transmittance at 1.8~{\textmu}m of our microscope with the objective lens attached is approximately 32\%, while the transmittance without an objective lens is approximately 60\%. Fluorescence photon signals were collected by the objective lens and detected by a photomultiplier tube (H7422-40p; Hamamatsu) placed after an emission filter (FF01-709/167 for TurboFP635, FF01-510/84 for Clover; Semrock).

% The pulses were sent into a microscope system consisting of 
% galvanometer scanners (Cambridge Technology), 
% a scan lens (SL50-3P, Thorlabs), 
% a trinocular tube (U-TR30IR, Olympus), 
% and a water-immersion objective (LUMPLFLN40XW, Olympus),
% as is described in \cite{OSAC3-1428}. %shown in Fig.~\ref{fig:microscope}.

% \begin{figure}[t]
%     \centering
%     \includegraphics[width=.5\textwidth]{setup_microscope_detail.pdf}
%     \caption{Optical setup within the microscope}
%     \label{fig:microscope}
% \end{figure}

% The galvanometer scanners were controlled with the ScanImage 
% software\cite{Pologruto_2003_BEO_2_13}.
% The total transmission of the optics within the microscope is 
% approximately 10\%.
% The major loss is due to the microscope objective, where the transmission is less than 30\%, 
% because there was no objective designed for 1.8~{\textmu}m laser 
% commercially available at the moment and we had to use an objective designed for
% another wavelength.
% % Signals were collected in the backward direction. 
% A photomultiplier tube (R3896, Hamamatsu) was used to detect 
% the fluorescence signal emitted from the samples
% in the backward direction.

\begin{figure}
    \centering
    \includegraphics[width=0.8\textwidth]{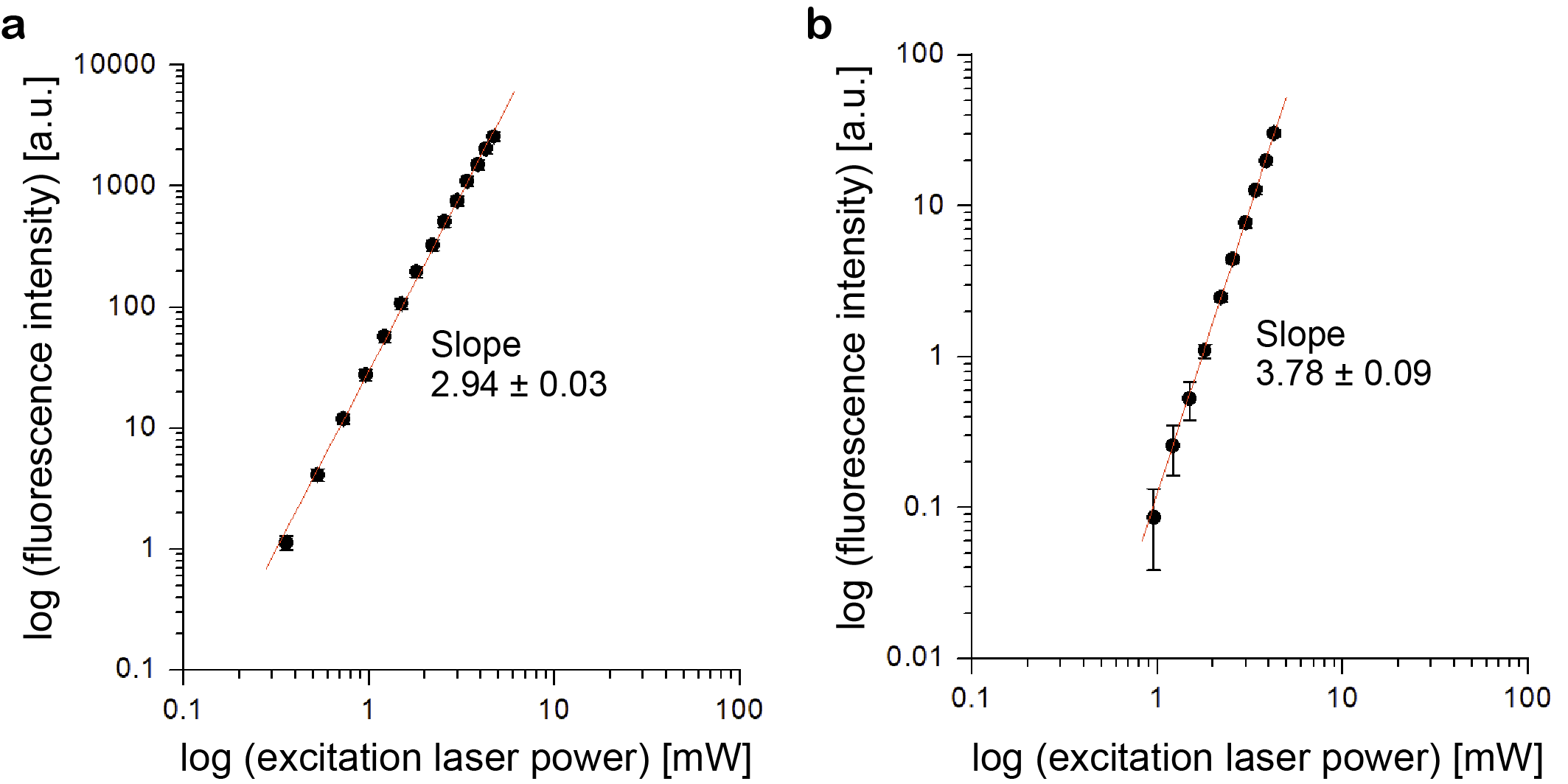}
    \caption{\label{fig:intensitydependence}Logarithmic plots of the dependence of (a) 3P, (b) 4P excited fluorescence intensity for TurboFP635 and Clover on excitation laser power. The slopes are indicated in each figure.}
\end{figure}

To test whether our setup can induce 3P excitation processes, 
we measured the dependence of the fluorescence from TurboFP635 (Katushka)~\cite{NMeth4-741} and Clover~\cite{NMeth9-1005} on the excitation laser power at 1.8~{\textmu}m (Fig.~\ref{fig:intensitydependence}). 
The purified fluorescent proteins from E.coli were dissolved in standard phosphate-buffered saline (PBS) at pH7.4. 
The fluorescent protein solution was sandwiched between two glass cover glasses and placed under the objective lens with D\textsubscript{2}O as immersion water. 
The slopes in the log-log plots confirmed 3P excitation for TurboFP635 (proportional to the cube) and 4P excitation for Clover (proportional to the fourth power).

\begin{figure}
  \centering
    \includegraphics[width=0.8\textwidth]{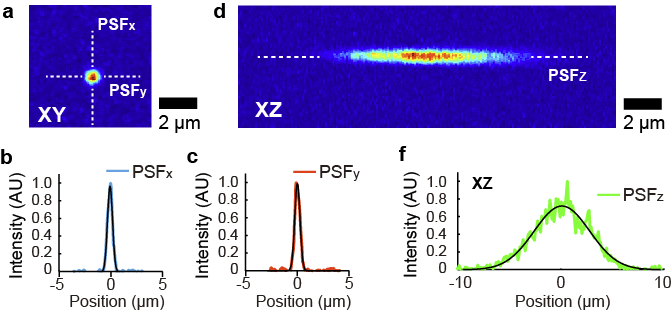}
  \caption{%Results of three-photon excitation measurements using the microscope. 
%  A typical image obtained by observing red fluorescent beads dispersed in agarose gel.
  Experimental measurement of the point spread function (PSF) with fluorescent beads. Images were acquired with a 25x/NA1.0 water immersion objective. (a) PSF in the XY plane. (b, c) Intensity profiles and fitted curves along the indicated cross-sections. (d) PSF in the z-direction. (f) An intensity profile of (d) and fitted curve (black line). The scale bars are indicated in the figures.}
  \label{fig:3photon}
\end{figure}

% \begin{figure}[t]
%     \centering
%     \caption{A typical image obtained by observing red fluorescent beads dispersed in agarose gel.}
%     \label{fig:3photon}
% \end{figure}

To determine the spatial resolution of the microscope with the 1.8 µm laser, we fixed red fluorescent beads (0.2 µm, FluoSpheres, crimson fluorescent 625/645) in agarose gel and observed them. A typical image is shown in Fig. 5. We fitted the spots with a gaussian to determine the spot size. Since the spot size is much larger than the beads, the spot size should reflect the point spread function of the microscope system. We determined the spatial resolution of the microscope by fitting with Gaussian curves (FWHM 0.56 µm to x-, y-axis, FWHM 6.2 µm to the z-axis, the average of 7 measurements).

\begin{figure}
  \centering
 \includegraphics[width=0.8\textwidth]{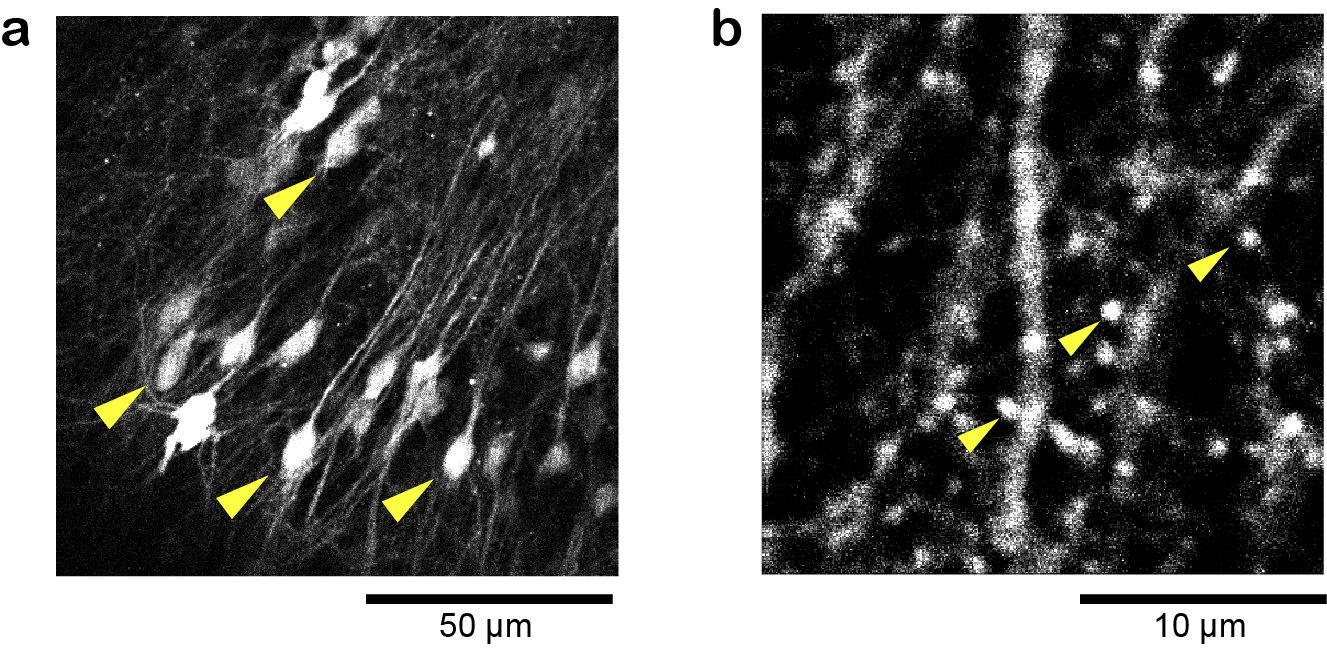}
  \caption{
Typical 3P fluorescence images of CA1 neurons expressing TurboFP635 in a hippocampal slice of a mouse brain. (a) Images of pyramidal neurons. Some of the somata are indicated by arrowheads. Scale bar, 50~{\textmu}m. (b) Images of dendrites and dendritic spines. Some of the spines are indicated by arrowheads. Scale bar, 10~{\textmu}m.
% Typical images of Neurons expressing TurboFP635 in hippocampal slices of rat brain.
  }
  \label{fig:cell-images}
\end{figure}

We next imaged a mouse hippocampal cultured slice with 3-photon excitation at 1.8~{\textmu}m. The hippocampus from a 6-day-old mouse was sliced into 350~{\textmu}m thickness, and 2 days later, the adeno-associated virus encoding CaMKII promoter-DIO-TurboFP635-WPRE3 in combination with low Cre expression~\cite{NCom12-751} was introduced into CA1 region at a depth of approximately 50~{\textmu}m from the surface of the slice. 
The slices were incubated at 35~$^\circ$C 5\% CO\textsubscript{2} for 17 days so that a sufficient level of TurboFP635 expression was obtained. 
For observation, the slice in PBS was taken out from the incubator and observed under the microscope. Figure~\ref{fig:cell-images} shows the typical image taken with the laser power of 8~mW at the sample. % (Fig.~\ref{fig:cell-images}).

% We further tested the microscope using living tissues from a rat brain.
% A hippocampus from a 6-day-old rat was sliced into 350~{\textmu}m thickness, 
% and the next day, the adeno-associated virus encoding TurboFP635 was introduced 
% at a depth of approximately 50~{\textmu}m from the surface of the slice.
% The slices were incubated at 35{\textcelsius} 5\% 
% CO\textsubscript{2} for 10 days so that a sufficient level of TurboFP635 expression was obtained.
% For observation, the slices were taken out from the incubator and 
% observed under the microscope within 1 hour so that the cells are still alive.
% Figure~\ref{fig:cell-images} shows some of the typical images
% taken with the laser power of 70~mW at the sample.
% The shapes of the neurons were clearly observed.

\begin{figure}
 %  \centering
 % \includegraphics[width=0.8\textwidth]{invivo3PE.png}
 %  \caption{(a) Mouse. (b) Experimental setup. (c) and (d) Neurons expressing TurboFP635 in the cortex of the mouse brain at the depths of 0.5~mm and 0.8~mm, respectively.}
    \centering
    \includegraphics[width=0.8\textwidth]{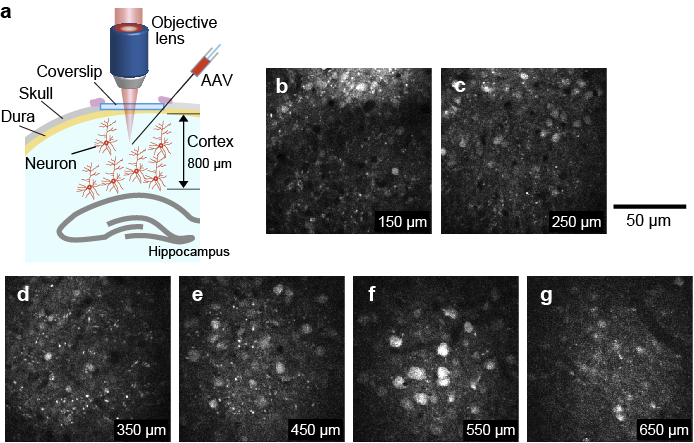}
    \caption{(a) Schematic drawing of {\itshape in vivo} imaging of neurons expressing TurboFP635. 
    TurboFP635 was expressed in cortical neurons of a mouse (C57BL/6N) and was excited by 3-photon excitation at 1.8~{\textmu}m. (b-g) Neurons expressing TurboFP635 at the depths of 150~{\textmu}m (b), 250~{\textmu}m (c), 350~{\textmu}m (d), 450~{\textmu}m (e), 550~{\textmu}m (f), and 650~{\textmu}m (g), respectively. Scale bar for x-y plane, 50~{\textmu}m. 
    }
\label{fig:invivo3PE}
\end{figure}

To demonstrate the capability of {\itshape in vivo} imaging by the 1.8~{\textmu}m laser, 
we performed 3-photon {\itshape in vivo} imaging of the mouse brain. 
To label excitatory neurons of the cortex with TurboFP636, adeno-associated virus (AAV) encoding CaMKII promoter-TurboFP635-WPRE3 (CaMKII promoter, TurboFP635, and WPRE3 were gifts from M. Ehlers, S. Sampson (Addgene plasmid \#78314)~\cite{microbiology162-966}, 
and BK. Kaang (Addgene plasmid \#61463)~\cite{MolecularBrain7-17}, respectively) was prepared as described previously~\cite{NCom12-751}. 
AAV was injected into the S1 region of the cortex at a depth of approximately 200, 400, 600, 800, 1000~{\textmu}m from the surface of the cortex as described previously~\cite{NCom12-751}. 
We typically waited about two weeks after AAV injection until a sufficient level of TurboFP635 expression was obtained. 
For the imaging, the cortical region in a mouse anesthetized with 1\% isoflurane was observed. 
We acquired a 700-µm-deep stack, taken with 2~{\textmu}m depth increments with ten frames averaging for single-frame acquisition (Fig.~\ref{fig:invivo3PE}). The average laser power required for 3-photon excitation at the surface and 800~{\textmu}m is $\sim$10 and $\sim$70~mW, respectively. During the image acquisition from 800 µm to the brain surface, the excitation laser powers were manually adjusted so that the image intensity was approximately the same. We successfully observed neurons at a depth of ~650 µm (Fig. 7d). The images were processed by ImageJ (National Institutes of Health; Bethesda, MD, USA).

% We further tested the microscope for in vivo imaging of a mouse brain.

\section{Four-photon (4P) microscopy}

\begin{figure}
  \centering
 \includegraphics[width=0.8\textwidth]{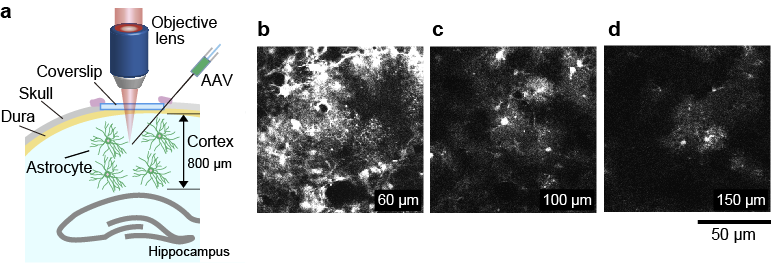}
  \caption{(a) Schematic drawing of {\itshape in vivo} imaging of astrocytes expressing Clover. 
  Clover was expressed in cortical astrocytes of a mouse (C57BL/6N) and was excited by 4-photon excitation at 1.8~{\textmu}m. (b-d) Astrocytes expressing Clover at the depths of 50~{\textmu}m (b), 100~{\textmu}m (c), and 150~{\textmu}m (d), respectively. Scale bar for x-y plane, 50~{\textmu}m.}
  \label{fig:invivo4PE}
\end{figure}

We also performed 4-photon {\itshape in vivo} imaging of astrocytes in the cortex at 1.8~{\textmu}m. 
To label astrocytes with Clover, adeno-associated virus (AAV) encoding gfaB1CD promoter-Clover-WPRE3 (gfaB1CD is a gift from B. Khakh, Addgene plasmid 44330) was injected into the S1 region of the cortex at a depth of approximately 200, 500, 1000~{\textmu}m from the surface of the cortex as described above. 
We acquired a 150-{\textmu}m-deep stack, taken with 2~{\textmu}m depth increments with ten frames averaging for single-frame acquisition (Fig.~\ref{fig:invivo4PE}). 
The average laser power required for 4-photon excitation at the surface and 150~{\textmu}m is $\sim$70 and $\sim$80~mW, respectively.
During the image acquisition from 150~{\textmu}m to the brain surface, the excitation laser powers were manually adjusted so that the image intensity was approximately the same. 
We successfully observed individual astrocytes at a depth of $\sim$150~{\textmu}m (Fig.~\ref{fig:invivo4PE}d).

\section{Conclusion}

We have successfully built a 1.8~{\textmu}m fiber laser system with commercially available standard components and lasers. There are no custom-designed components in the system. 
The pulse duration becomes shorter than 150~fs with a pulse energy of $\sim$1~{\textmu}J. 
We applied the laser for 3P and 4P microscopy and successfully observed TurboFP635-expressing cortical neurons at 0.7~mm depth with 3-photon excitation and Clover-expressing astrocytes at 0.15~mm depth with 4-photon excitation.

The performance of the laser system is not as high as the optical parametric amplifier pumped by several tens of watt ytterbium fiber lasers. However, our system costs much less than such laser systems. 
In addition, our system does not require delay-sensitive parts, i.e., all amplifications are on a single line. Another benefit of the system is the scalability of the repetition rate. 
While optical parametric amplifiers require several hundred watts of power to achieve repetition rates above 10~MHz, 
our system only needs to change the frequency of the pulse picker and increase the power of the cw laser pumping the final amplifier. 
It is much easier than increasing the power of the femtosecond pump laser for the optical parametric amplifier. 
This feature would be crucial to realize photon counting based fluorescence lifetime imaging by 3- or 4-photon excitation. 
We show that the pulse with a duration of 150~fs is enough for {\itshape in vivo} 3P and 4P microscopy. Thus, we believe that our system can be an alternative to the optical parametric amplifier system. 

% We have built a 1.8~{\textmu}m fiber laser system consisting of commercially available standard components and lasers. 
% There are no custom designed components in the system. 
% The pulse duration becomes shorter than 150~fs with a pulse energy of $\sim$1~{\textmu}J. We have applied the laser for three and four-photon microscopy and observed neurons 0.8~mm and 0.4~mm deep inside the cortex of a living mouse, respectively. 

% The performance of the laser system is not as high as 
% the optical parametric amplifier pumped 
% by a several tens of watt ytterbium fiber laser.  
% However, our system costs much less than such laser systems. 
% Unlike the optical parametric amplifier system, 
% our system does not include delay sensitive parts, 
% all amplifications are on a single line.  
% We are showing 
% that the pulse with the duration of 150~fs 
% is enough for in vivo three- and four-photon microscopy.  
% We believe that the system can be an alternative 
% to the optical parametric amplifier system. 

% Several optical parametric amplifier also provide 
% 1.3~{\textmu}m pulses 
% and the pulses were also used for two- and three-photon fluoresence imaging. 

% \section{Backmatter}

% Backmatter sections should be listed in the order Funding/Acknowledgment/Disclosures/Data Availability Statement/Supplemental Document section. An example of backmatter with each of these sections included is shown below.

\begin{backmatter}
\bmsection{Funding}
%SENTAN (JST, Japan Science and Technology), Consortium for Photon Science and Technology (JST), the Joint Studies Program (2016-2018) of the Institute for Molecular Science, 
Core Research for Evolutional Science and Technology
(CREST) (JPMJCR17N5), 
Frontier Photonic Sciences Project of National Institutes of Natural Sciences (NINS), 
The Naito Foundation, 
MEXT/JSPS KAKENHI (22H02724, 22H05549, 21H05703, JP22H04926).

\bmsection{Acknowledgments}
The authors would like to thank Kazuhiko Ogawa of FiberLabs Inc., who provided the ZBLAN fibers used for the oscillator and fiber amplifier. 
In addition, they gave us many insightful comments and suggestions. 

\bmsection{Disclosures}
The authors declare no conflicts of interest.

\bmsection{Data availability} Data underlying the results presented in this paper are not publicly available at this time but may be obtained from the authors upon reasonable request.

% \bmsection{Supplemental document}
% See Supplement 1 for supporting content. 

\end{backmatter}

\bibliography{references}

%%%%%%%%%% If preparing manually:
% \newcommand{\enquote}[1]{``#1''}

% \bibitem{Zhang:14}
% Y.~Zhang, S.~Qiao, L.~Sun, Q.~W. Shi, W.~Huang, L.~Li, and Z.~Yang,
%   \enquote{Photoinduced active terahertz metamaterials with nanostructured
%   vanadium dioxide film deposited by sol-gel method,}
%   {\protect\JournalTitle{Optics Express}} \textbf{22}, 11070--11078 (2014).

% {Optical Society}, \enquote{{OSA Publishing},}
%   \url{http://www.osapublishing.org}.

% \bibitem{FORSTER2007}
% P.~Forster, V.~Ramaswamy, P.~Artaxo, T.~Bernsten, R.~Betts, D.~Fahey,
%   J.~Haywood, J.~Lean, D.~Lowe, G.~Myhre, J.~Nganga, R.~Prinn, G.~Raga,
%   M.~Schulz, and R.~V. Dorland, \enquote{Changes in atmospheric consituents and
%   in radiative forcing,} in \enquote{Climate Change 2007: The Physical Science
%   Basis. Contribution of Working Group 1 to the Fourth assesment report of
%   Intergovernmental Panel on Climate Change,}  S.~Solomon, D.~Qin, M.~Manning,
%   Z.~Chen, M.~Marquis, K.~B. Averyt, M.~Tignor, and H.~L. Miler, eds.
%   (Cambridge University Press, 2007).

% \end{thebibliography}

\end{document}